\documentstyle[aps]{revtex}

\begin{document}

\begin{flushright}
{\large\bf $<<$Journal of Physics: Condensed Matter 1999
\\ Vol. 11, Issue: 29, Pages: L349-354 $>>$}
\end{flushright}

\centerline{\Large {\bf New Boundary Bound States in an Open }} \centerline
{\Large {\bf Quantum Spin Chain}} \centerline{\bf Zhan-Ning Hu} 
\centerline{\ Institute of Physics and Center for Condensed Matter
Physics,} 
\centerline{\ Chinese Academy of Sciences, Beijing 100080, China}

\begin{center}
\begin{minipage}{5in}
\centerline{\large\bf   Abstract}
New boundary bound states (BBS) are found of an integrable model with 
the magnetic impurities located at the edges of an open Heisenberg  
spin chain. These bound states carry the real energy 
and are formed by three or five imaginary modes of the rapidities. 
These imaginary modes of the rapidities give the non-zero antisymmetric 
wave functions and the moments of the centers of the bound 
states are zero. It means that these bound states are arisen by 
the magnetic impurities and localized at the edges of the correlated 
system. The Kondo screening occurs for the antiferromagnetic 
spin chain with the ferromagnetic impurities-electrons 
exchange interaction.

\smallskip

\end{minipage}
\end{center}

%\newpage

Magnetic impurities in one-dimensional (1D) strongly correlated electron
systems or quantum spin chains have been the focus of intense
investigation.
These strong correlated systems can be described in term of Luttinger
liquid 
\cite{FN67} and the behavior of the impurities in the 1D quantum systems is
rather different from that in a Fermi liquid\cite{cho10,cho11}. The
availability of nonperturbative techniques allows us to have detailed
pictures understanding the relevant physics and revealing some very
interesting phenomena such as Kondo problem\cite{koon,cho10} and pinning of
bound states in the low dimensional strong correlated electron systems and
quantum spin chains. Experimentally, the magnetic impurities implant in
carbon nanotubes or quantum wires and analogical phenomena, for example,
the 
$x-$ray boundary effects, metal point-contact spectroscopies, etc, renew
also the investigations of these problem.

The quantum inverse scattering method (QISM) and the Bethe ansatz (BA)
techniques have provided us the very effective tools to study the magnetic
impurities in the 1D quantum system. These methods have been used
successfully to deal with the Kondo impurity in a free electron host, the
magnetic impurities in spin chains and the mixed valent behavior of
hybridization (Anderson-like, with hybridized impurity and host wave
functions) \cite{zj2,zj3euop,euroro1}. Recently, in correlated electron
hosts, the properties of the magnetic impurities have been studied in a
series of very interesting papers\cite{zj40Foe,zj41}. The {\em periodic}
boundary conditions were imposed on the electron host and spin chains for
all these cases. Kane and Fisher investigated a 1D repulsive interacting
system in the presence of a potential barrier and pointed out that it
corresponds to a chain disconnected at the barrier site at low energy
scales 
\cite{cho3}. This can be effectively described by the {\em open} boundary
conditions which are well studied by the boundary conformal field
theory\cite
{cho} and the BA methods\cite{cho57,zj5}. Zvyagin get that the low-energy
magnetic behaviors of an impurity in a chain with periodic boundary
conditions and with open boundary conditions coincide up to mesoscopic
corrections of order of $L^{-1}$ with $L$ being the length of the
system\cite
{zj41}. Now we know that several methods have been used for introducing the
impurities into the integrable models of correlated electrons and quantum
spin chains with the {\em open} boundary conditions. The first one for the
construction of the impurity models is dependent on the main idea that the
spectral parameters in the scattering matrix of the model rely on the
difference of the particles' rapidities. This method has been firstly used
to the impurity model of the 1D quantum lattice gases with the periodic
conditions by Eckle, Punnoose and R$\stackrel{..}{o}$mer\cite{euroro1},
where the integrable condition continues to be satisfied under an arbitrary
local shift of the related parameter (See, for example\cite{hupuprb}). The
second method depends on that the scattering matrix of the bulk of the
$t-J$
model relies on the tangent ( or cotangent ) functions of the half moments
of the electrons\cite{ijmp}. The very interesting thing is that the
corresponding Hamiltonian with the magnetic impurities has a simple and
compact form\cite{hujpa}, which is different from the one in the open
Hubbard impurity model\cite{huhue}. The boundary scattering matrix between
the impurity and the electron can be factorized as two terms: one is
similar
as the $R$ matrix and another is similar as the inverse of the $R$ matrix
with the inverse spectral parameter, justly as rightly obtained by Zvyagin
and Johannesson in their very interesting work\cite{hidde}. They revealed
the existence of a hidden Kondo effect driven by forward electron
scattering
off the impurity related to this property of the boundary scattering
matrix.

As is well known, the bound states can be formed for the strongly
correlated
electron system and quantum spin chains within the charge or spin sectors
and they are very important to decide the thermodynamics of the system and
the its low temperature properties. Despite the success of the QISM and BA
approach to the investigation of the magnetic impurities in correlated
system, it is not very clear that the impurities contribute to the bound
states of the strongly correlated system with the {\em open }boundary
conditions. In this letter, we discuss this problem in details.

In general a bound state can be formed by several complex rapidities of the
particles, such as charges or spins. The total energy of these complex
modes
and the total moments of the ones should be real. Under the open boundary
conditions, the rapidities $u_j$ of the charges or spins should satisfy
also
that $u_j\neq \pm u_l$ when $j\neq l.$ Otherwise, the wave function is
zero,
which means that this situation is forbidden. When the complex rapidities
form a bound state, the above three conditions should be satisfied for the
correlated system under the open boundary conditions. For clarity we here
focus on the case of the quantum Heisenberg model. When the two impurities
with the arbitrary spins are coupled to this open quantum spin chain, the
Hamiltonian of the system can be written as 
\begin{equation}
H=\frac J2\sum_{j=1}^{N-1}{\vec{\sigma}}_j\cdot {\vec{\sigma}}_{j+1}+J_L{\ 
\vec{\sigma}}_1\cdot {\vec{S}}_L+J_R{\vec{\sigma}}_N\cdot {\vec{S}}_R,
\end{equation}
where ${\vec{\sigma}}_j$ are the Pauli matrices; ${\vec{S}}_{L,R}$ are the
impurity moments with the arbitrary spins $S_{L,R}$. The site number of the
bulk is $N$. $J_{L,R}$ are two arbitrary real constants which describe the
coupling between the bulk and the impurities and can be parameterized as 
\begin{equation}
J_{L,R}=\frac J{\left( S_{L,R}+\frac 12\right) ^2-c_{L,R}^2}  \label{tra01}
\end{equation}
with the arbitrary constants $c_{L,R}.$ This Hamiltonian can be
diagonalized
by using the standard Bethe ansatz scheme. The eigenvalue of the energy of
this impurity system is 
\[
E(\lambda _1,\lambda _2,\cdots ,\lambda _M)=\sum_{j=1}^M\frac{-J}{\lambda
_j^2+\frac 14}+\sum_{l=L,R}J_lS_l+\frac{J(N-1)}2, 
\]
with the following Bethe ansatz equations: 
\[
\left( \frac{\lambda _j+\frac i2}{\lambda _j-\frac i2}\right)
^{2N}\prod_{l=L,R}\prod_{r=\pm 1}\frac{\lambda 
_j+i\left( S_l+rc_l\right) }{
\lambda _j-i\left( S_l+rc_l\right) }\qquad 
\]
\begin{equation}
\qquad =\prod_{l=1(l\neq j)}^M\prod_{r=\pm 1}\frac{\lambda _j+r\lambda 
_l+i}{
\lambda _j+r\lambda _l-i}.  \label{be001}
\end{equation}
Now we study the boundary bound states in detail for the above Heisenberg
impurity model. From definition (\ref{tra01}) we know that $-c_{L,R}$ is
equivalent to $c_{L,R}$ because they give the same Hamiltonian (1). So,
without losing generality, we restrict the parameters $c_{L,R}$ to
non-negative values in the following discussion.

When the coupling in the bulk is antiferromagnetic $(J>0)$ and the coupling
between the bulk and the impurity is ferromagnetic, the system has the
following boundary bound states (BBS): 
\begin{eqnarray}
\lambda _{3,1} &=&i(S_{L,R}-c_{L,R}),  \label{5504} \\
\lambda _{3,2} &=&-\frac i2(S_{L,R}-c_{L,R}-1),  \label{5505} \\
\lambda _{3,3} &=&-\frac i2(S_{L,R}-c_{L,R}+1),  \label{5506}
\end{eqnarray}
when the parameters $c_{L,R}$ in the regime
$S_{L,R}+1/2<c_{L,R}<S_{L,R}+1$.
These two BBS, one BBS for each ends ( $L$ and $R$) of the chain, are
formed
by the three imaginary modes of $\lambda $, respectively. They carry the
energy 
\begin{equation}
E_{L,R}^{(3)}=\frac{12J\left[ 3(S_{L,R}-c_{L,R})^2-2\right] }{\left[
4(S_{L,R}-c_{L,R})^2-1\right] \left[ (S_{L,R}-c_{L,R})^2-4\right] }.
\label{e502}
\end{equation}
The moments of the centers of the BBS are $\sum_{j=1}^3\lambda _{3,j}=0.$
It
means that this kind of bound states is localized at the edges of the
system. Of course, the energy and the moments of the centers of the BBS are
all real and the spin rapidities satisfy that $\lambda _{3,1}\neq \pm
\lambda _{3,2}\neq \lambda _{3,3}\neq \pm \lambda _{3,1}$ which ensure that
the antisymmetric wave functions of the system are not zero. Therefore,
these BBS satisfy all the physical demand. By making the transformations $
\lambda _{3,j}\rightarrow -\lambda _{3,j}$ $(j=1,2,3)$ in the relations
(\ref
{5504}-\ref{5506}), the BBS with the three imaginary modes can be obtained
also for the Heisenberg impurity model with
$S_{L,R}+1/2<c_{L,R}<S_{L,R}+1$.
They carry the energy also as the form (\ref{e502}).

When the coupling in the bulk is antiferromagnetic $(J>0)$ and the coupling
between the bulk and the impurity falls also into the antiferromagnetic
regime, the above BBS (\ref{5504}-\ref{5506}) with the three imaginary
modes
are formed under the condition $1/3+S_{L,R}<c_{L,R}<1/2+S_{L,R}$. The
corresponding imaginary modes with the transformations $\lambda
_{3,j}\rightarrow -\lambda _{3,j}$ $(j=1,2,3)$ are also the BBS in this
case. The energy expression (\ref{e502}) do not change the forms and the
moments of the centers of the bound states are also zero. By taking the
logarithm of the Bethe ansatz equation (\ref{be001})and introducing the
distribution functions of the spin rapidities, we can get the integral
equations of the impurity model for the ground state. Then we have that the
self-magnetization of the ground state is $S_L+S_R-1$ for the two up
impurity spins, or $1-S_L-S_R$ for the two down impurity spins, or $\pm
\left( S_L-S_R\right) $ for the one up and one down impurity spins. The
similar procedure gives that the self-magnetization of the ground state is
also $\pm \left( S_L-S_R\right) $ when the coupling in the bulk is
antiferromagnetic but the coupling between the bulk and the impurity is
ferromagnetic. When the parameters $c_{L,R}$, which describe the coupling
between the impurities and the bulk, satisfy that $
S_{L,R}<c_{L,R}<S_{L,R}+1/3,$ there are the BBS formed by the three
imaginary modes as relations (\ref{5504}-\ref{5506}) and the ones by the
transformations $\lambda _{3,j}\rightarrow -\lambda _{3,j}$ $(j=1,2,3).$
They carry the energy (\ref{e502}) and have the zero moments of the centers
of the BBS. In this case, the coupling between the bulk and the impurities
are in the antiferromagnetic regime if the exchange interaction in the bulk
is antiferromagnetic, too. Furthermore, in this case ($
S_{L,R}<c_{L,R}<S_{L,R}+1/3$), there are two bound states (one for each
ends
of the chain) formed by the following five imaginary modes of $\lambda ,$
respectively, 
\begin{eqnarray}
\lambda _{5,1} &=&i(S_{L,R}-c_{L,R}),  \label{xx01} \\
\lambda _{5,2} &=&i(S_{L,R}-c_{L,R}+1),  \label{xx02} \\
\lambda _{5,3} &=&i(S_{L,R}-c_{L,R}-1),  \label{xx03} \\
\lambda _{5,4} &=&-\frac i2(3S_{L,R}-3c_{L,R}-1),  \label{xx04} \\
\lambda _{5,5} &=&-\frac i2(3S_{L,R}-3c_{L,R}+1).  \label{xx05}
\end{eqnarray}
They carry the energy 
\begin{equation}
E_{L,R}^{(5)}=\frac{20J\left[ 7(S_{L,R}-c_{L,R})^2-6\right] }{\left[
9(S_{L,R}-c_{L,R})^2-4\right] \left[ 4(S_{L,R}-c_{L,R})^2-9\right] }.
\label{x504}
\end{equation}
The moments of the centers of above bound states formed by the five
imaginary modes of $\lambda $ are $\sum_{j=1,2,\cdots ,5}\lambda _{5,j}=0.$
So they localize at the two edges of the Heisenberg spin chain and $\lambda
_{5,j}\neq \pm \lambda _{5,l}$ if $j\neq l$ $(j,$ $l=1,2,\cdots ,5)$ which
ensure that the system has a non-zero antisymmetric wave function.
Similarly
as the case of the BBS of the three imaginary modes, the transformations $
\lambda _{5,j}\rightarrow -\lambda _{5,j}$ $(j=1,2,\cdots ,5)$ give also
the
boundary bound states of the system and they do not change the expression
of
the energy and the moments of the centers of the bound states are also
zero.
By the use of the method mentioned above, we get that the
self-magnetization
of the model is same as the situation $1/3+S_{L,R}<c_{L,R}<1/2+S_{L,R}.$
The
above BBS satisfy the three conditions of the complex modes.

In the following part, we describe simply the self-magnetization and the
BBS
of the ferromagnetic Heisenberg impurity model. By solving the Bethe ansatz
equations in the thermodynamic limit, we get that the self-magnetization of
the system is $S_L+S_R+N/2$ when the coupling between the impurities and
the
bulk is also ferromagnetic. Otherwise, the self-magnetization is $
-S_L-S_R+N/2$ for the antiferromagnetic exchange interaction between the
impurities and the bulk. When one impurity has the ferromagnetic
interaction
with the bulk and another impurity has the antiferromagnetic interaction
with the bulk, the self-magnetization of the system has the form $\pm
\left(
S_L-S_R\right) +N/2.$ When $S_{L,R}+1/2<c_{L,R}<S_{L,R}+1,$ the system has
the BBS formed by the three imaginary modes (\ref{5504}-\ref{5506})and the
BBS carry the energy (\ref{e502}). By making the transformations $\lambda
_{3,j}\rightarrow -\lambda _{3,j}$ $(j=1,2,3),$ the corresponding bound
states are also the BBS of the system. They satisfy all of the three
conditions and the coupling between the bulk and the impurities is
antiferromagnetic. When the coupling between the bulk and the impurities is
ferromagnetic, the system has the BBS which can be formed by the three
imaginary modes (\ref{5504}-\ref{5506}) or the five imaginary modes (\ref
{xx01}-\ref{xx05}) with $S_{L,R}<c_{L,R}<S_{L,R}+1/3$. They carry the
energies as expressions (\ref{e502}) and (\ref{x504}), respectively. Of
course, the corresponding imaginary modes with the transformations $\lambda
_{3,j}=-\lambda _{3,j}$ $(j=1,2,3)$ and $\lambda _{5,l}=-\lambda _{5,l}$ $
(l=1,2,\cdots ,5)$ form also the BBS of the impurity model. When $
1/3+S_L<c_L<S_L+1/2,$ we have the BBS (\ref{5504}-\ref{5506}) or the
inverse
of rapidities and the exchange interaction between the bulk and the
impurities is ferromagnetic. Notice that the system may have the other form
of the impurity bound states in the above restricted range of the impurity
couplings $J_{L,R}$. It is also an open problem to find out the impurity
bound states for other values of the impurity couplings for the strongly
correlated system.

The above discussion shows that the ferromagnetic and antiferromagnetic
Heisenberg spin chains with the magnetic impurities have always the BBS
arisen by the impurities when the strengths of the interactions between the
bulk and the impurities are proper. These bound states carry the real
energy
and the moments of the centers of the BBS are zero. They satisfy all of the
three conditions of the imaginary modes. For the ferromagnetic Heisenberg
model, the system has the BBS\ formed by the three imaginary modes and the
five imaginary modes of the spin rapidities when the coupling between the
bulk and the impurities is ferromagnetic. When the coupling between the
bulk
and the impurities is antiferromagnetic, the system has only the BBS
contained the three imaginary modes of the rapidities. For the
antiferromagnetic Heisenberg model, the system has the BBS\ formed by the
three imaginary modes and the five imaginary modes of the spin rapidities
when the coupling between the bulk and the impurities is antiferromagnetic.
When the coupling between the bulk and the impurities is ferromagnetic, the
system has only the BBS contained the three imaginary modes of the
rapidities. And the Kondo screening -as predicted by Furusaki and Nagaosa 
\cite{cho10}- exists for the antiferromagnetic Heisenberg model with the
ferromagnetic coupling between the bulk and the impurities. At zero
impurity
couplings ($J_{L,R}=0$), the system has impurity bound states, although
trivial, which just correspond to the $2S_{L,R}+1$ spin states of the
impurity. By turning on the boundary coupling, the BBS can be formed by the
imaginary modes of the rapidities and the number of the BBS might change.
The BBS can affect the ground state of the whole system when the boundary
coupling is strong enough, which is under investigation. New properties of
the specific heat, excited state, dressed energy, $et$ $al$ can be
introduced due to the impurity couplings. We point out that the similar BBS
with three or five imaginary modes contributed by the magnetic impurities
can be found also for the strongly correlated electron systems such as the
Hubbard model and the $t-J$ model with the open boundary conditions in the
charge sectors. The BBS carry the energy and satisfy the above three
conditions of the imaginary modes. Finally, the way the string affects the
distribution of the rapidities is also the interesting subject for the
further investigation.

To conclude, we have found that the BBS arisen by the magnetic impurities
in
a correlated host under the open boundary conditions. These BBS are formed
by three or five imaginary modes such as charges or spins. The imaginary
modes of the bound states due to the magnetic impurities satisfy that: (i)
the total energy is real; (ii) the total moment of the imaginary modes is
real (zero); (iii) the absolute values of the imaginary modes are
different.
These BBS carry the energy and pin at the edges of the system. And the
Kondo
screening exists for the antiferromagnetic Heisenberg model with the
ferromagnetic coupling between the bulk and the impurities.

%\newpage

\end{document}